\documentclass[a4paper,11pt]{article}
\usepackage{jinstpub} 

\proceeding{Light Detection In Noble Elements - LIDINE 2025\\
21--24 October 2025\\
Hong Kong}

\title{\boldmath A compact Optical Liquid Argon Facility at Roma Tre }

\author[a,b,1]{H.~Shi\note{Corresponding author.},}
\author[c]{V.~D'Andrea,}
\author[a,b]{G.~Salamanna,}
\author[c]{K.~Szczepaniec,}
\author[b]{D.~Tagnani}
\affiliation[a]{Department of Mathematics and Physics, Universit\`{a} degli Studi Roma Tre,\\
Rome, Italy}
\affiliation[b]{Istituto Nazionale di Fisica Nucleare, Sez. di Roma Tre,\\
Rome, Italy}
\affiliation[c]{INFN - Laboratori Nazionali del Gran Sasso, Assergi, Italy}

\emailAdd{hexi.shi@uniroma3.it}

\abstract{
In this paper we present a compact test facility for the measurement of optical properties of liquid argon as scintillator. 
The setup is under preparation at Roma Tre and it has a volume of 40~L liquid argon, 
which is liquefied from argon gas with a purity of $\ge 99.9999\%$ vol.
To readout the scintillation photons from liquid argon with the highest intensity near 127~nm, 
we use the vacuum ultraviolet silicon photomultipliers from Hamamatsu. 
By submerging the photon detectors directly inside the liquid argon, 
we can eliminate the systematics from the wavelength shifter and light guides 
which have been commonly used to detect the scintillation photons of liquid argon. 
}


\arxivnumber{arXiv:2601.10363} 

\begin{document}
\maketitle
\flushbottom


\section{Introduction}
\label{sec:intro}
Liquid argon (LAr) has been used in neutrino and dark matter experiments 
as an active medium because of its excellent properties in charge yield and transport, 
as well as its capacity as a scintillator~\cite{Heindl2010}. 
To detect LAr scintillation photons of the vacuum ultraviolet (VUV) wavelength range, 
standard method uses wavelength shifting materials to produce secondary photons 
in the visible wavelength range to which typical photomultipliers have the sensitivity. 
The additional optical processes and photon paths contribute to the 
systematic uncertainties which are difficult to quantitatively evaluate 
in the measurements for the light yield and attenuation length~\cite{Ishida1997, Jones2013, Calvo2018}.

In recent years Hamamatsu has developed the VUV4-series windowless Silicon Photomultipliers (SiPMs), 
with an appreciable Photon Detection Efficiency (PDE) about 10--20\%~\cite{Hamamatsu, Pershing22} 
down to the 127--128~nm peak wavelength of the argon scintillation photons.
Without the wavelength shifting material or the light guide, 
improvement in the systematic uncertainties is expected in the measurement of the optical properties of LAr. 
The direct detection of the VUV photons will also enable new solutions and detector design
.

Within the framework of the LEGEND-200 experiment~\cite{L-200}, 
the LLAMA detector has successfully used the Hamamatsu VUV SiPMs inside the 
64~m${^3}$ LAr passive shielding volume in the LEGEND-200 to monitor the impurity level~\cite{Schwarz21}. 

At Roma Tre University, we are preparing the Optical Liquid Argon Facility (OLAF),  
with a cylindrical volume of 40~L LAr to host VUV SiPMs and photon sources inside.
The first of the two main goals of the facility is 
to characterize the optical properties of LAr as a scintillation detector 
to be used in the LEGEND-1000 experiment~\cite{L-1000} as active shielding. 
Secondly, the facility is expected to function as a test bench for the R\&D of the LAr scintillation detector in general. 
The compact size of the OLAF setup is important for a fast turnaround time to test different configurations and designs of the measurement.
In the next section, we will introduce the experimental setup in detail.

\section{Experimental setup}
\label{sec:setup}
\begin{figure}[htbp]
\centering
\includegraphics[width=.80\textwidth]{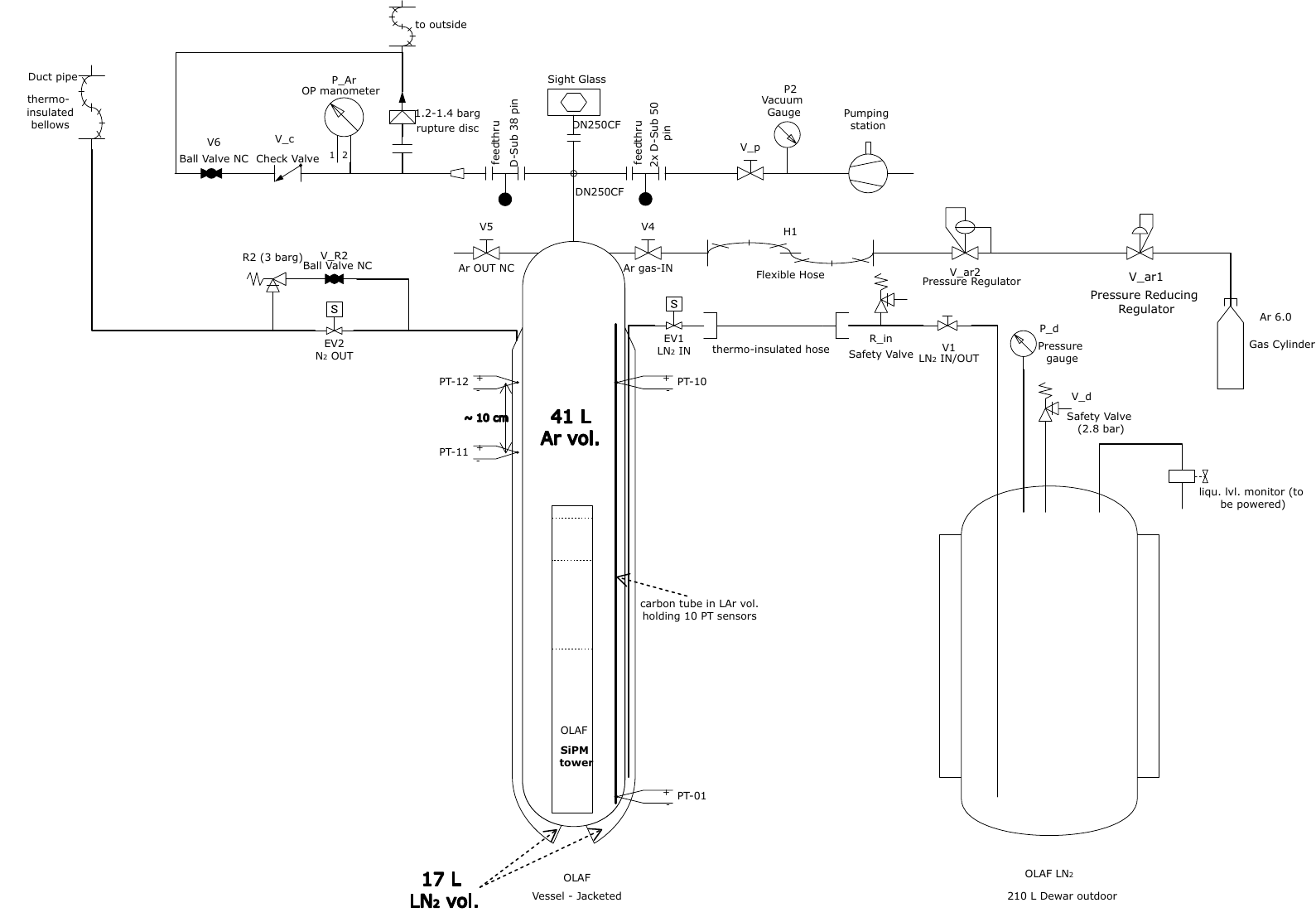}
\caption{The P\&ID diagram of the OLAF setup. 
Details in text.\label{fig:pid}}
\end{figure}

The OLAF setup consists of two main parts, to be introduced in the two subsections that follow. 
The first part is the cryogenic system to liquefy Ar 6.0 gas
~\footnote{Argon 6.0 is a term for argon gas with a purity level of $\ge 99.9999\%$ vol, the highest available commercially.} 
and to maintain LAr during measurement using liquid nitrogen (LN$_2$) as coolant. 
The second part is the readout of the vacuum ultraviolet (VUV) SiPMs. 
The schematic of the setup is indicated in the P\&ID diagram in figure~\ref{fig:pid}. 

\subsection{Cryogenics for LAr}
\label{subsec:cryo}
\begin{figure}[htbp]
\centering
\includegraphics[width=.8\textwidth]{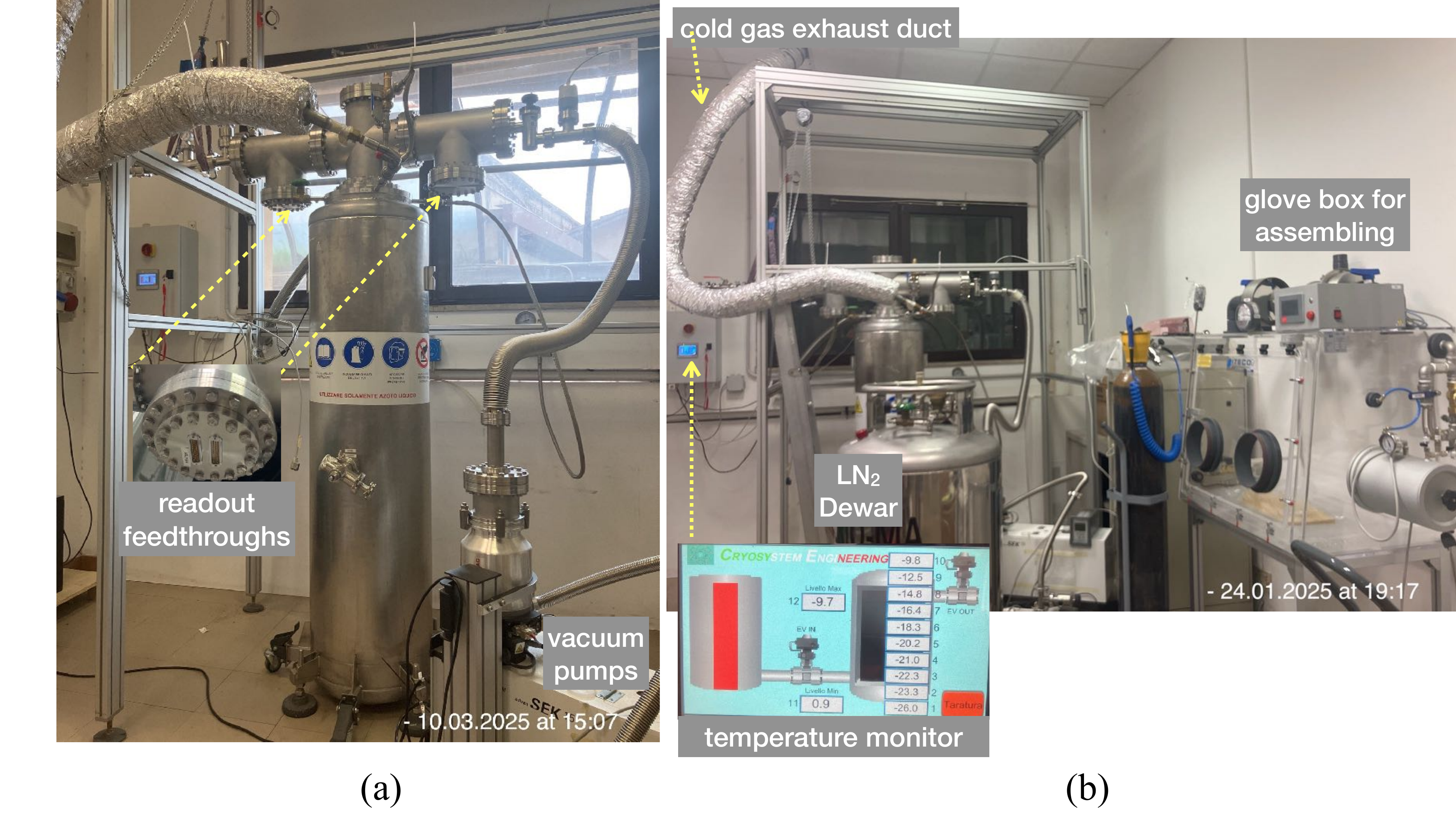}
\caption{The OLAF setup up at the laboratory of Roma Tre: 
(a) the main vessel with the LAr surrounded by LN$_2$ and vacuum jacketed vessels; 
(b) photo showing other auxiliary components for OLAF operation. 
\label{fig:setup}}
\end{figure}

The central part of the cryogenic system is a cylindrical vessel made of stainless steel, 
about 2~m high, as shown in figure~\ref{fig:setup}. 
The innermost vessel can host LAr up to 40~L, 
surrounded by a jacketed volume of about 20~L where the LN$_2$ is filled. 
An external vacuum layer then thermally isolates the inner volumes from the surroundings. 
The three layers are welded together at the top part of the vessels, 
as shown in the photo in figure~\ref{fig:setup}a. 
A 200~L Dewar is used to continuously provide LN$_2$ to OLAF. 
The thermally insulated pipes were equipped, as shown in figure~\ref{fig:setup}b, 
to vent the cold gas to the roof outside the laboratory. 
The top of the LAr volume is fitted with a transparent sight glass through which 
the interior can be visually inspected. 
Temperature sensors placed in multiple vertical locations indicate the liquid level 
of the LAr and LN$_2$ volumes. 
The feed-through flanges as shown in figure~\ref{fig:setup}a provide separate sockets 
for the SiPMs and the temperature sensors. 

\paragraph{Liquefaction of argon}
When the LN$_2$ volume of OLAF is filled, 
the thermal contact of the jacketed wall provides a cooling power of a few kilowatts to the argon volume. 
By controlling the flow of the argon gas, 
we can produce tens of liters of LAr in a few hours~\footnote{This includes the time required to cool the vessels from room temperature with LN$_2$}. 
At 1~atm, the boiling temperature of LN$_2$ (77~K) is lower than the freezing temperature of LAr (83~K). 
To prevent LAr from  freezing, 
we regulate the pressure of the gas phase of the volume of LN$_2$ with a relief valve that vents at 2.5--3~bar ("R2" in figure~\ref{fig:pid}). 
In stable conditions, the nitrogen volume is sealed by two electric valves indicated by EV1 and EV2 in figure~\ref{fig:pid}, 
and the boiling temperature of LN$_2$ stays above 86 K. 
As the upper rims of the LAr and LN$_2$ vessels have direct thermal contact with the flange at the top of the OLAF setup, 
a heating power of a few tens of Watts will warm the inner vessels, 
causing the LAr to evaporate constantly.
For measurements longer than a week, 
we need to replenish LN$_2$ and Ar 6.0 gas to produce more LAr.

\subsection{Photon detector layout and readout}
\label{subsec:elec}
\begin{figure}[htbp]
\centering
\includegraphics[width=.75\textwidth]{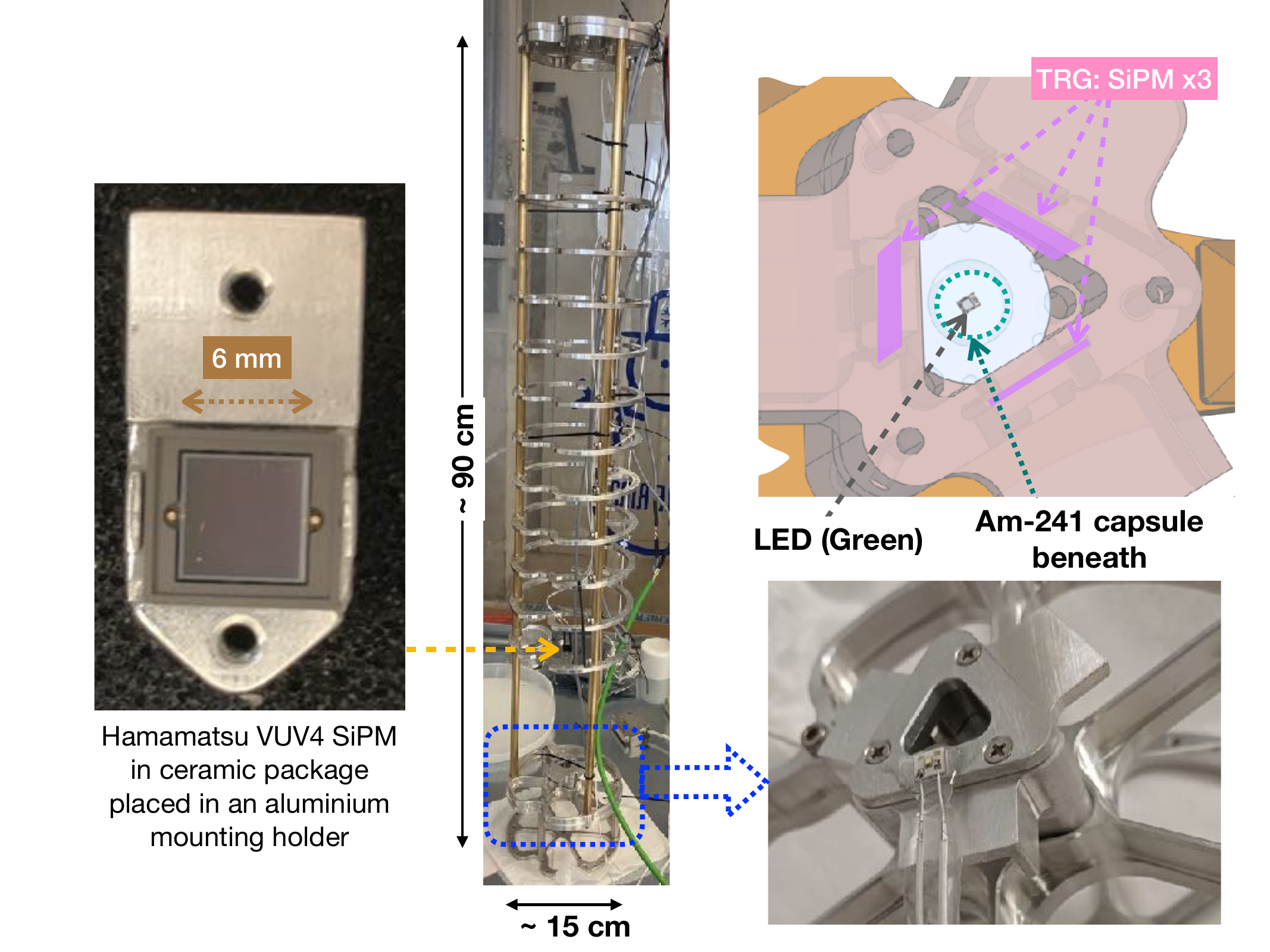}
\caption{Middle: the prototype mechanical tower structure which hosts the SiPMs and the photon sources; 
left: metallic holder to host a single SiPM(in ceramic package by Hamamatsu) and to fix it to the tower; 
top right: a 3D rendering that shows the radioactive source housing and the three SiPMs defining the trigger;  
bottom right: prototype to test the mounting of LED source to the source holder.
\label{fig:photon_detection}}
\end{figure}

The layout of the SiPMs and the light source structure was designed 
with the primary goal of measuring the attenuation length and the yield of the LAr scintillation photons. 
The prototype of the mechanical structure for mounting the SiPMs and the photon source 
is shown in figure~\ref{fig:photon_detection}. 
The exterior diameter and the height of the cylindrical tower shown in the photo 
are adapted to the dimension of the OLAF LAr volume inner diameter 
and the attenuation length of the LAr scintillation photon of the order of some tens of centimeters. 
The tower has multiple rings each holding one SiPM, 
placed between 15~cm and 80~cm from the photon sources at the bottom layer of the tower. 
In figure~\ref{fig:photon_detection}, the photo on the left shows a prototype holder 
to host the SiPM (delivered by Hamamatsu with a ceramic package~\footnote{Specifically the S13370 series with package designed for cryogenic applications~\cite{Hamamatsu}.}) and to be fixed to a ring. 
The SiPMs face downward towards the photon sources, 
and their positions on the rings are staggered to avoid shadowing the SiPMs on the upper levels of the tower. 

\paragraph{Photon source and calibration}
To generate liquid argon scintillation photons of known intensity, 
we will use a fully encapsulated Am-241 source. 
The $\alpha$ particles from the primary alpha decay are shielded by the capsule, 
and only the accompanying 59.5~keV $\gamma$ (35.9\% of the time) propagates in LAr  
with a mean free path of less than 1.5~cm.
Using the coincidence of three SiPMs near the point where the $\gamma$ vanishes, 
one can obtain a well-defined trigger for the position and time of the generated scintillation photons. 
The 3D rendering in the top-right of figure~\ref{fig:photon_detection} illustrates 
the layout for the scintillation photon trigger. 
We will also place a green LED on the source holder for preliminary function validation  
and timing calibration of the SiPMs. 
The figure~\ref{fig:photon_detection} bottom-right photo shows 
the LED glued with optical cement~\footnote{EJ-500 two-component optical cement from ELJEN Technology.} to the prototype holder, 
before the structure was completely submerged in LAr for testing.

\paragraph{Readout electronics}
The power supply and the amplification of the SiPM output are realized with 
a front-end board developed by Roma Tre for the LEGEND-200 experiment~\cite{AbrittaCosta2023}. 
The module is packaged in the form of a single NIM unit, 
and each unit can be connected to 12 SiPMs. 
Coaxial cables of 5~m long are used to connect the SiPMs via a multi-pin feed-through to the front-end board which operates at room temperature. 
The bias voltage for each connected SiPM can be set separately, 
while a universal amplification rate of 40 applies to all SiPM signals. 
The differential outputs of the front-end board are then connected to 
a CAEN V2740~\cite{caen_v2740} digitizer module, 
which has 64 input channels with a sampling rate of 125 MS/sec. 
This commercial  digitizer has an FPGA on board 
and can operate with multiple firmwares provided by CAEN. 
A common feature for all the firmwares is the capability to 
define a trigger logic during the run-time between the digitized inputs. 
This feature allows the triple coincidence trigger logic for 
the scintillation photons to be defined within the digitizer. 
When the LED is used as the photon source, 
the digitizer can take an external trigger, 
which is synchronized with the digital signal modulating the power supply to the LED. 
In addition to digitizing the waveform of the front-end board's output, 
the pulse shape discrimination firmware also provides the integrated charge values of the input 
waveform with two different user-defined timing gates. 
Using these two integrated charge values, 
we can apply the standard technique of pulse shape discrimination 
for particle type identification for fast online analysis.

\paragraph{Data acquisition}
We have implemented the data acquisition of a single V2740 digitizer unit 
with the open-source package Maximum Integrated Data Acquisition System (MIDAS)~\cite{midas} 
developed at PSI and TRIUMF~\footnote{Paul Scherrer Institute, Switzerland; TRIUMF, Canada}. 
The program is compiled on an Alma Linux operating system and uses 
the C++ library provided by CAEN to communicate with the digitizer.

\section{Current status and plans}
\label{sec:status}
\begin{figure}[htbp]
\centering
\includegraphics[width=.5\textwidth]{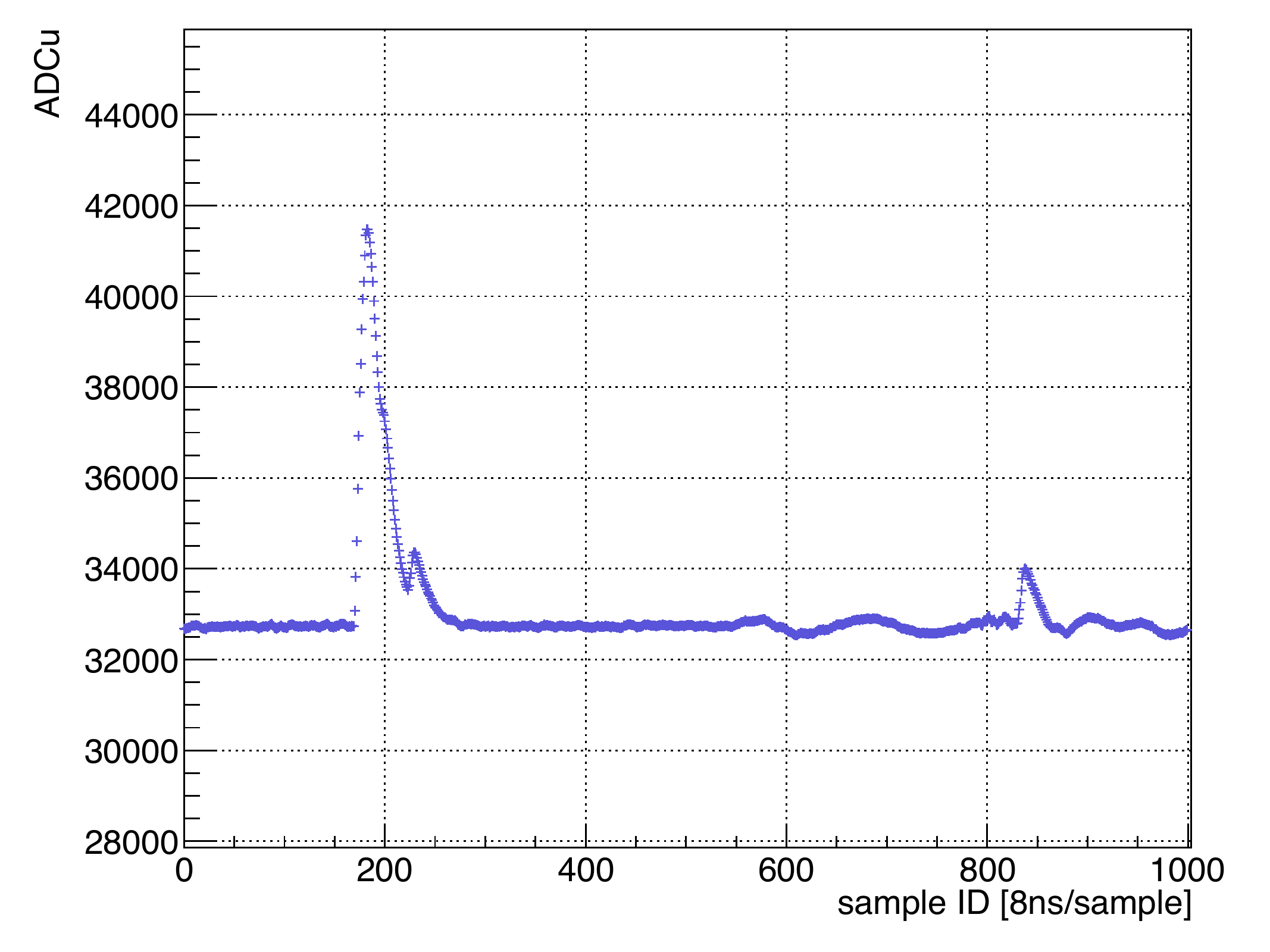}
\caption{A typical waveform of the LAr scintillation photon detected by the VUV SiPM. 
The digitizer operated at the sampling rate of 8~ns per sample; 
the vertical axis is in ADC unit. \label{fig:waveform}}
\end{figure}

By the end of 2025, 
we have tested the complete readout chain for a single SiPM, 
which was mounted to the prototype tower structure and submerged in LAr. 
The mechanical mounting and gluing of the LED was also tested in LAr. 
Digitizer data were taken with the single SiPM 
triggered by the scintillation photons of LAr from environmental radiation.  
Figure~\ref{fig:waveform} shows a typical digitized waveform 
when the LAr scintillation photon is detected by the VUV SiPM in OLAF.  
The polarity and the baseline level settings of the digitizer are not essential in this context. 
This result demonstrates that 
we have succeeded to observe the time structure and amplitude of the waveform, 
which are shown in this 8~$\mu$s time window (8~ns per sample). 
Refined analysis on the timing and the number of detected photons will follow.

In the next steps, 
we will implement the readout for multiple SiPMs 
and the Am-241 source holder structure with the SiPMs for the triple-coincidence trigger, 
then we will test all SiPMs and photon sources in LAr. 
The noise level measurement and the optimization of the bias voltage for the SiPM 
in LAr will follow, 
as well as the validation of the performance of the trigger. 
The final production of the mechanical tower structure in oxygen-free copper 
shall be delivered in 2026, 
and it will replace the prototype structure. 
After validating the readout for all the SiPMs and the trigger performance, 
we will start the physics measurement on the optical properties of the LAr.

\acknowledgments
In preparing the design of the OLAF setup, we thank the LLAMA team at Technical University of Munich: 
Patrick Krause, Laszlo Papp, Stefan Schönert,  Christoph Vogl, and Mario Schwarz. 
Special thanks to the mechanical workshop team at INFN Sez. Roma Tre: 
Gianfranco Paruzza, Gabriele Paruzza, and Alessandro Top, 
who produced the prototype SiPM tower structure and provided invaluable advice 
on the final design for commercial production. 
We also thank the HAMMER team at INFN Sez. Roma1 for the 3D printing service.



\end{document}